# μSR and neutron diffraction studies on the tuning of spin-glass phase in partially ordered double perovskite SrMn$_{1-x}$W$_x$O$_3$

Poonam Yadav[1], Shivani Sharma[1], Peter J.Baker[2], Pabitra K. Biswas[2], Ivan da Silva[2], and Niranjan P. Lalla[1]

[1]UGC-DAE Consortium for Scientific research, Indore-452001, India
[2]ISIS Facility, Rutherford Appleton Laboratory, Chilton, Didcot OX11 0QX, United Kingdom

## Abstract

Tunability of the partially ordered double perovskite (PODP) and coexisting spin-glass phase in SrMn$_{1-x}$W$_x$O$_3$ (x=0.20 to 0.40) have been studied using neutron powder diffraction (NPD), muon spin relaxation (μSR), and magnetic susceptibility (χ) measurements. Structural studies reveal that SrMn$_{1-x}$W$_x$O$_3$ undergoes a quasi-continuous transformation from simple perovskite (*Pm-3m*) to PODP (*P2$_1$/n*) phase as x increases. $\chi_{dc}(T)$ and $\chi_{ac}(T)$ measurements show a sharp cusp-like peak at a spin glass transition, T$_g$. The muon relaxation rate (λ) peaks at T$_g$ following a critical growth, given by $\lambda=\lambda(0).t^{-w}$ *[t=(T-T$_g$)/T$_g$]*. However, no long-range magnetic order is observed in NPD below T$_g$. These measurements confirm a tunable spin-glass freezing in SrMn$_{1-x}$W$_x$O$_3$ with T$_g$ monotonously decreasing with W content, which we attribute to tuning the relative concentration of the coexisting Mn-O-Mn and Mn-O-W-O-Mn anti-ferromagnetic super-exchange pathways altering the geometric frustration.

## 1. Introduction

The occurrence of exotic magnetic ground states like spin-liquid [1-5], valence bond glass [6], spin-glass [7-11], and spin-ice [12,13] have been of immense theoretical and experimental interest for several decades [14-17]. These exotic magnetic phases originate from geometric frustration arising from competing nearest neighbor (NN) and next nearest neighbor (NNN) antiferromagnetic (AFM) interactions between moments arranged on triangular, tetrahedral [18], Kagome [19] and Shastry-Sutherland [20] type lattices. Due to the presence of tetrahedral topology in its structure, as shown in the inset of Fig.1 (a), magnetic double perovskites (DPs) [6,21-23] $A_2BB'O_6$ have been extensively studied regarding such exotic magnetic ground states. In magnetic DPs [24] B/B′ (diamagnetic/magnetic) cations follow a tetrahedral topology, therefore an AFM super-exchange interaction cannot be simultaneously satisfied at each lattice site, and hence long-range AFM order gets suppressed. Such an interaction gives rise to cooperative phenomena resulting in frustration-driven freezing of spin configurations, such as valence bond glass $Ba_2YMoO_6$ [6] and spin-glass as in $Sr_2CaReO_6$ [21], $Ba_2YReO_6$ [22] and $Sr_2MgReO_6$ [23]. However, there are examples, where despite the presence of tetrahedral topology, magnetic DPs undergo normal AFM ordering [25,26].

An ideal DP has *Fm-3m* structure, but it is often distorted due to steric pressure and temperature variations [27,28] resulting in its structural variants like $P4_2/n$, *Pnma*, $P2_1/m$, $P2_1/n$ etc. The b-site chemical ordering of B/B′ cations still exists [28-30], but the tetrahedral topology can be distorted relaxing the condition of frustration and resulting in AFM ordering. In such DPs the NN AFM interaction is an extended super-exchange, active along B′-O-B-O-B′ pathways and leads to type-I AFM, e.g. $Ca_2LaRuO_6$ [31], $Sr_2YRuO_6$ [32], $Sr_2LuRuO_6$, $Ba_2YRuO_6$, $Ba_2LuRuO_6$ [33], $La_2LiRuO_6$ [34] and $Sr_2TeMnO_6$ [35]. It has been shown that DPs having diamagnetic atoms like W [25, 26,36,37], Nb [38] or Mo [26], at its b-sites, exhibit type-II AFM in which the NNN spins are AFM aligned.

It is worth noting that in the case of perovskites the frustrated magnetic ground states have been studied mostly on perfectly ordered DPs. No attempts have been made to study the evolution of magnetic-frustration as a function of b-site substitution. The flexibility of the perovskite structures for b-site substitution provides a unique opportunity for such studies. For DP structures it is not yet known, under the favorable conditions of ionic-size and charge-states of B/B′, what minimum substitution can result in the formation of $A_2B'_{2-x}B_xO_6$ type partially ordered double perovskites (PODP) or whether a PODP even forms. The physical properties of perovskites are mostly governed by the $BO_6$ octahedra. Hence the quasi-continuous variation in PODPs can provide tunability of the properties. The coexistence of B′-O-B′ and B′-O-B-O-B′, mixed super-exchange pathways in a PODP, should alter the magnetic interactions and result in a frustrated spin-lattices with exotic magnetic phases.

To address these questions we have carried out studies on a distorted DP with composition typified by $A_2B'_{2-x}B_xO_6$ taking $Sr_2Mn_{2-x}W_xO_6$ as a case study. $Sr_2MnWO_6$ is a perfectly ordered distorted DP with G-type AFM order. We found that PODP forms for x ≥ 0.3. Compositions with x < 0.3 remain simple disordered perovskites. Also, the magnetic and structural properties of PODPs $Sr_2Mn_{2-x}W_xO_6$ are entirely different from the fully ordered $Sr_2MnWO_6$ [25,26,37,38]. In place of long-range AFM order [25,26], we found a spin-glass state with $T_g$ systematically decreasing with x. Such tunability of the site occupancy will be common to nearly all DPs, and hence this approach can prove a novel way for tailor-making the properties of other DPs as well.

2. **Experimental**

$Sr(Mn_{1-x}W_x)O_3$ compounds with x = 0.20 to 0.40 were prepared following the conventional solid-state reaction route using 99.99% pure $SrCO_3$, $MnO_2$ and $WO_3$. The thoroughly ground stoichiometric mixture of the ingredients was calcined and sintered respectively at 1200°C for 24 h and 1450°C for 12 h. Phase purity characterization was done using powder X-ray diffraction. Magnetization-vs.-temperature (M-T) measurements were carried out under zero-field cooled (ZFC) and 500 Oe field-cooled (FC) conditions using SQUID-VSM (QD). To investigate the microscopic nature of the magnetic ground states, neutron diffraction (NPD) (200-5K), using the GEM instrument, and muon spin relaxation (μSR) measurements in zero-field (ZF) and 0 to 3000 Oe longitudinal-fields (LF) (125-2K), using the MuSR instrument, were performed at the ISIS Pulsed Neutron and Muon Source, U.K.

3. **Results**

**3.1 Neutron diffraction studies**

In literature, the structure of $Sr_2Mn^{4c}W^{4d}O_6$ has been attributed to two possible space-groups $P4_2/n$ [25] and $P2_1/n$ [26]. We refined the room temperature NPD data of all the studied PODPs based on *P2₁/n* using **Jana-2006** [39]. Figure 1(a) presents a typical nuclear structure refined NPD data for SMWO30 using *P2₁/n*. The refinement of PODP structures was done following coupled *4c* and *4d* occupancies described as $Sr_2(Mn_{1-x}W_x)(Mn_yW_{1-y})O_6$. For details on the refinement see supplementary materials. Figure 1(b,c) shows typical example of the refined crystal structure of PODP phase. A systematic variation in the bias of partial occupancies of Mn *(4c)* and W *(4d)* has been clearly realized. The variations in the partial occupancies of Mn and W and the Mn-O-Mn bond angles, along *c*-axis and in the *a-b* plane, with increasing x, is shown in Table 1 of the supplementary materials. The refined structural data also shows that with increasing W the bending of Mn-O-Mn bond along *c*-axis and the second type of Mn-O-Mn bond in the *a-b* plane increases with x. The bending of Mn-O-Mn is likely to cause canting of the moments [40,41].

### 3.2 Magnetic studies

Temperature dependent dc magnetic susceptibility ($\chi_{dc}$-*vs-T*) of $Sr(Mn_{1-x}W_x)O_3$ were measured for x= 0.20 to 0.40. Figures 2 (a) and (b) show a typical ZFC, and 500 Oe FC $\chi_{dc}$-*vs-T* for x= 0.20 and 0.30 respectively. The ZFC $\chi_{dc}$-*vs-T* shows a sharp cusp-like feature at $T_g$ that systematically shifts from 53 to 14 K as x increases from 0.20 to 0.40. The variation of $T_g$ and the corresponding $\chi_{dc}(T_g)$ is shown in the inset of Fig.2(a). The cusp-like feature has also been reported by Q. Lin *et al.* [42] for the fully ordered DP of $Sr_2MnWO_6$. ZFC cusp shows bifurcation with FC curve. The dc-susceptibility increases with W. To explore the dynamic behavior of the short-range spin-correlations, ac-susceptibility ($\chi_{ac}$-*vs-T*) measurements were carried out at frequencies ranging from 31 to 937 Hz. The $\chi_{ac}$-*vs-T* data for SMWO20 and SMWO30 is shown in Fig.2 (c-d). The $\chi_{ac}$-*vs-T* also shows a cusp-like peak, the frequency dependence of which is shown in the insets to each panel.

To check for the occurrence of any long or short ranged magnetic order, NPD measurements were performed at temperatures below the cusp-like anomalies. Figure 3 shows the comparison of the low-q (high-d) regions of the low-temperature NPD profiles of SMWO30, SMWO32 and SMWO40. In literature occurrence of a pronounced AFM ordering peak at ~ 9.2 Å is reported for $Sr_2MnWO_6$ by Azad et al. [25] and Munoz et al. [26], but no such feature was evident in our data, so we find no evidence for long-ranged AFM ordering.

**3.3 µSR measurements**

To investigate the spin-glass phase in the PODPs phase of SrMn$_{1-x}$W$_x$O$_3$ muon spin relaxation (µSR) measurements were carried out on Sr(Mn$_{0.80}$W$_{0.20}$)O$_3$, Sr(Mn$_{0.70}$W$_{0.30}$)O$_3$ and Sr(Mn$_{0.60}$W$_{0.40}$)O$_3$ using 100% spin-polarized pulsed µ$^+$-beam (duration 70ns, period 20ms) at low-temperatures (125- 1.5K) at the ISIS facility, UK [43]. µSR probes much wider time domain, $10^5$ to $10^{10}$ Hz, of relaxation than $\chi_{ac}(T)$ [44]. The asymmetry A(t)-vs-time (t) was measured from 0.1 to 32µs in zero-field (ZF) and longitudinal-fields (LF) from 0 to 0.3 Tesla. The ZF A(t)-vs-t µSR spectra collected down to 1.5 K is presented in the supplementary materials. The nature of decay of the asymmetry function $A(t)$ appears to follow an exponential form as

$$A(t) = A_0.\exp(-\lambda.t) + A_{bg} \qquad (1)$$

where $A_0$ and $\lambda$ are the initial asymmetry and the relaxation rate respectively with a flat background $A_{bg}$. The A(t)-vs-t µSR spectra of SMWO20, SMWO30 and SMWO40 show qualitatively similar behavior, except for lowering of the transition temperature. $A_0$ and $\lambda$ were determined by fitting the A(t) data with eqn. (1) using Mantid software [45]. Figures 4(a-f) show the temperature dependence of $A_0$, $\lambda$ and $A_{bg}$ as obtained from the fit of the A(t) data of the above samples. In each panel of Fig. 4(a-f) we could identify three different regimes marked as (I), (II) and (III). The flat behavior of $A_0$ and $\lambda$ in regime (I) shows the paramagnetic nature of the samples. As the temperature is lowered across the characteristic temperature ($T_g$), in region (II), $A_0$ drops sharply. The steep drop in $A_0$ indicates that the muon spins are depolarized well within the pulse duration of 70 ns. The corresponding peaks in $\lambda$ in regime (II) show that the time scale of the fluctuation of local-field is slowing down and crosses the µSR time window of 10 ps to 1µs, around $T_g$. Both these anomalies show correspondence with the *ac/dc* $\chi(T)$ data. Figure 5(a) shows the *log($\lambda$) - vs - log{(T-$T_g$)/$T_g$}* plots for the three studied samples. The linear behavior of this plot indicates that as T approaches $T_g$ the muon relaxation rate grows critically i.e. $\lambda=\lambda(0).r^{-w}$, $r=\{(T-T_g)/T_g\}$. The critical growth of $\lambda(T)$ in turn indicates critical slowing-down of the spin-fluctuations. The complete absence of long-range magnetic order in NPD and the cusp-like peak in *ac/dc* $\chi(T)$ indicate that the slowing down of the spin-fluctuation is due to spin-glass transition at $T_g$.

To distinguish between the cases of fluctuation or freezing dominance, we collected µSR spectrum for SMWO30 at various longitudinal fields at 5 K. Figure

5(b) shows the *A(t)-vs-t* data collected from 0 to 3000 Oe. Since we are unable to decouple the muon spins from the internal magnetic fields they experience, we can conclude that the internal fields are significantly larger than the largest field used of 0.3 T.

## 4. Discussion

The bifurcation in ZFC and FC of $\chi_{dc}$-*vs-T* below the cusp, the frequency-dispersion of the peak maxima in $\chi_{ac}$-*vs-T*, the complete absence of the signature of any magnetic order in NPD and the critical growth of muon relaxation-rate $\lambda(T)$ i.e. $\lambda=\lambda(0).r^{-w}$ in zero-field μSR studies, clearly reveal that $Sr(Mn_{1-x}W_x)O_3$ for x= 0.20 to 0.40 undergoes a spin-glass transition. The significantly low value of magnetization, even at fields as high as 7T, is due to inherent moment frustration in the spin-glass state. The opening of the M-H loop (see supplementary materials) below $T_g$ is due to field irreversibility characteristics of spin-glass [15-17] which is seen even for canonical spin-glass systems too [46-48]. Minor increase in magnetization and the susceptibility is due to the weakening of the AFM interaction on doping of non-magnetic atom W at the Mn site. On W doping the effective magnetic moment also decreases. The spin-glass transition temperature $T_g$ is also supposed to decrease with decreasing moment concentration [49-51], in agreement with our observations.

In $SrMnO_3$ all Mn are in 4+ charge state. Tungsten (W), being in 6+ charge state, causes enhanced conversion of $Mn^{4+}$ into $Mn^{3+}$ while substituted in $Sr(Mn_{1-x}W_x)O_3$. Presence of 4+/3+ mixed valence of Mn gives rise to a series of interesting charge and spin ordered states [52] till x ≤ 0.18. As confirmed through soft x-ray absorption studies [53], for x≥ 0.16, besides $Mn^{4+}$ and $Mn^{3+}$, $Mn^{2+}$ also starts appearing in $Sr(Mn_{1-x}W_x)O_3$, which increasingly dominates over the $Mn^{3+}$/ $Mn^{4+}$ concentration. As x approaches 0.5, $Mn^{2+}$ maximizes and $Mn^{3+}$/ $Mn^{4+}$ goes to zero. Recently, Wang *et al.* [54] have also reported similar changes in the charge state of Mn on W substitution in $Sr(Mn_{1-x}W_x)O_3$. $SrMnO_3$, i.e. x=0 case, is a G-type AFM [55, 56] with NN $Mn^{4+}$ $t_{2g}$ moments interacting anti-ferromagnetically through $Mn^{4+}$-$O^{2-}$- $Mn^{4+}$ super-exchange pathway. At the other extreme of x=0.5,i.e. in the case of fully ordered DP $Sr_2MnWO_6$, the $Mn^{2+}$ cations are also ordered G-type anti-ferromagnetically [25] but now through $Mn^{2+}$-$O^{2-}$-$W^{6+}$-$O^{2-}$-$Mn^{2+}$ extended super-exchange pathway. Thus at the two extremes of the substitution levels, the interaction between NN Mn moments is AFM. While NN moments are AFM aligned the NNN moments will be FM aligned. The situation of intermediate substitutions, say, e.g. x =

0.20 to 0.40, can be represented as in Fig. 6(a), which shows a finite size Mn-O plane (2D) of PODP $Sr(Mn_{1-x}W_x)O_3$. In this finite Mn-O plane if we consider just a single row, e.g. the one highlighted with green dotted-line, the NN Mn moments, e.g. M1 and M2 interact anti-ferromagnetically but the interaction between M1 and M3, which should have been FM now will be AFM due to the presence of intermediate $W^{6+}$ in the extended super-exchange pathway $Mn^{2+}-O^{2-}-W^{6+}-O^{2-}-Mn^{2+}$, i.e. substitution of $W^{6+}$ flips the M3 moment anti-ferromagnetically *w.r.t.* M2. Had the system been 1D, then in the present case of NN super-exchange interaction, flipping of M3 and thereby all other moments in the left side (along with the green dotted-line), would have been possible without any energy cost, avoiding moment frustration in 1D. But when the system is 2D or 3D, as the present case of magnetic DP, flipping of M3 will independently try to reorient all the other moments around it. For example if we consider such an interaction in 2D, say along the path highlighted through red dotted-line in Fig.6(a), we find that flipping of M3 demands flipping of the very starting moment M1 itself, which is physically impossible and therefore in 2D or 3D AFM systems the moments will get geometrically frustrated leading to a spin-glass phase.

In $Sr(Mn_{1-x}W_x)O_3$, at a lower substitution level of $x < 0.16$, there are mostly $Mn^{4+}$ and $Mn^{3+}$, and we observed a variety of magnetically ordered CO-AFM phases [52]. With increasing concentration of W the $Mn^{+3}$, which is necessary for charge/orbital order, decreases. Hence the long-range charge/orbital ordered phase will get suppressed. Even in such situations magnetic frustration, as described in the above, may arise giving rise to spin-glass phase. The spin-glass phase observed for SMWO20 is a similar case. In PODP $Sr(Mn_{1-x}W_x)O_3$ the effective NN AFM interactions, arising due to of Mn-O-Mn and Mn-O-W-O-Mn super exchange pathways, gets continuously optimized with x. Due to the smaller Mn-Mn distance in the Mn-O-Mn superexchange pathway, the corresponding AFM interaction will be larger than that of the Mn-O-W-O-Mn. Thus with increasing W the Mn-Mn NN AFM interaction continuously decreases. Therefore the frustration and hence the spin-glass freezing temperature ($T_g$) also continuously decreases. That is what has been observed. The initiation of 3D percolation of Mn-O-W-O-Mn pathways will give rise to short-range PODPs, which will finally grow as long-range ordered DP. Our electron diffraction studies show that the limit is x = 0.25, see supplementary materials. In Fig. 6(b) we summarize the phase-diagram of $Sr(Mn_{1-x}W_x)O_3$, showing various magneto/structural phases as a function of W substitution[52].

## 5. Conclusion

Based on the above-described studies we conclude that on W substitution at Mn site the G-type anti-ferromagnetic $SrMnO_3$ undergoes a transformation to simple perovskite spin-glass phase at x ~ 0.20 and then partially ordered double perovskite (PODP) spin-glass phases with x ≥ 0.3. The spin-glass transition temperature $T_g$ continuously decreases in the range of 55 to 15 K with increasing W. The spin-glass phase arise due to magnetic frustration originating from constraints imposed by coexisting Mn-O-Mn and Mn-O-W-O-Mn extended super-exchange pathways with 2D and 3D AFM interactions in the PODP phase. The local field in spin-glass phase is more than 0.3 T. The spin-glass freezing is due to $Mn^{+2}$ cations. A magnetic phase-diagram has been presented. We finally conclude that under otherwise favorable b-site doping condition, a magnetic perovskite phase can be continuously transformed to PODP phase with tunability of the related physical properties, e.g. spin-glass phase in the present study.

**Acknowledgements:** Authors gratefully acknowledge the Director Dr. A. K. Sinha and the Centre-Director Dr. V. Ganesan for their constant support and encouragement. Discussion with Dr. A. Banerjee is sincerely acknowledged. Dr. A. Banerjee and Dr. R. J. Choudhary are acknowledged for magnetic measurements. Experiments at the ISIS Neutron and Muon Source were supported by a beam time allocation from the Science and Technology Facilities Council (U.K.).

**Figure captions:**

**Fig.1.** (a) Rietveld refined time of flight neutron powder diffraction profile of $Sr(Mn_{1-x}W_x)O_3$ for x = 0.30. Inset shows the structure of a double perovskite phase, highlighting the presence of tetrahedral topology of B/B' cations. (b) A typical crystal structure model of a partially ordered double perovskite of $Sr(Mn_{1-x}W_x)O_3$ obtained after Rietveld refinement. (c) The biased Mn occupancy at 4c and W occupancy at 4d sites for x=0.30. Atoms have been indicated in different colors.

**Fig.2.** (a-b) The temperature variation of $\chi_{dc}(T)$ for $Sr(Mn_{1-x}W_x)O_3$ respectively for x=0.2 and 0.3. Presence of cusp-like peak can be clearly seen. Inset shows variation of spin-glass transition temperature $T_g$ and $\chi_{dc}(T_g)$ as a function of W substitution. (c-d) $\chi_{ac}(T)$ variation of $Sr(Mn_{1-x}W_x)O_3$, for x = 0.20 and 0.30. Frequency dispersion of the cusp-like peaks can be clearly seen in the insets.

**Fig.3.** Enlarged view of the low-q (high-d) regions of bank-2 TOF-NPD profiles of SMWO30, SMWO32 and SMWO40. Unlike Ref. [26] these profiles clearly depict the absence of any peak at ~ 9.2 Å (indicated by dashed vertical line), showing AFM ordering in the DP phase.

**Fig.4.** The temperature variation of $A_0$, $\lambda$ and $A_{bg}$ as obtained from the Mantid fit of the A(t)-vs-t data. It can be seen that the initial asymmetry $A_0$ decreases and $\lambda$ peaks at $T_g$ of the corresponding samples.

**Fig.5.** (a) $log(\lambda)$ - vs - $log\{(T-T_g)/T_g\}$ plot showing critical behavior of the spin-freezing across glass-transition. (b) Longitudinal fields dependence of the A(t)-vs-t for SMWO30 at 5K taken from zero to 3000 Oe.

**Fig.6.** (a) Proposition for the possible cause of frustration in a perovskite structure having extended super-exchange path ways. The blue colored moments show the usual moments with orientations guided by Mn-O-Mn superexchange interaction, whereas the magenta colored moments are the flipped view of the same moments but now guided by Mn-O-W-O-Mn extended super-exchange interaction (b) Magnetic phase-diagram of $Sr(Mn_{1-x}W_x)O_3$ as a function of W substitution [51].

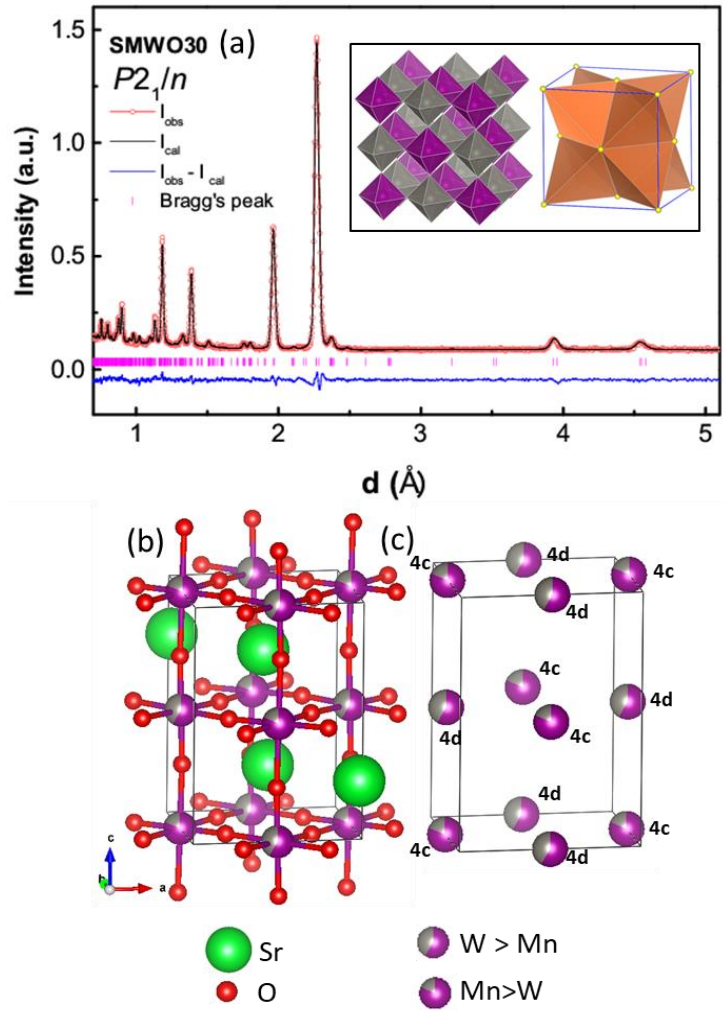

**Fig.1**

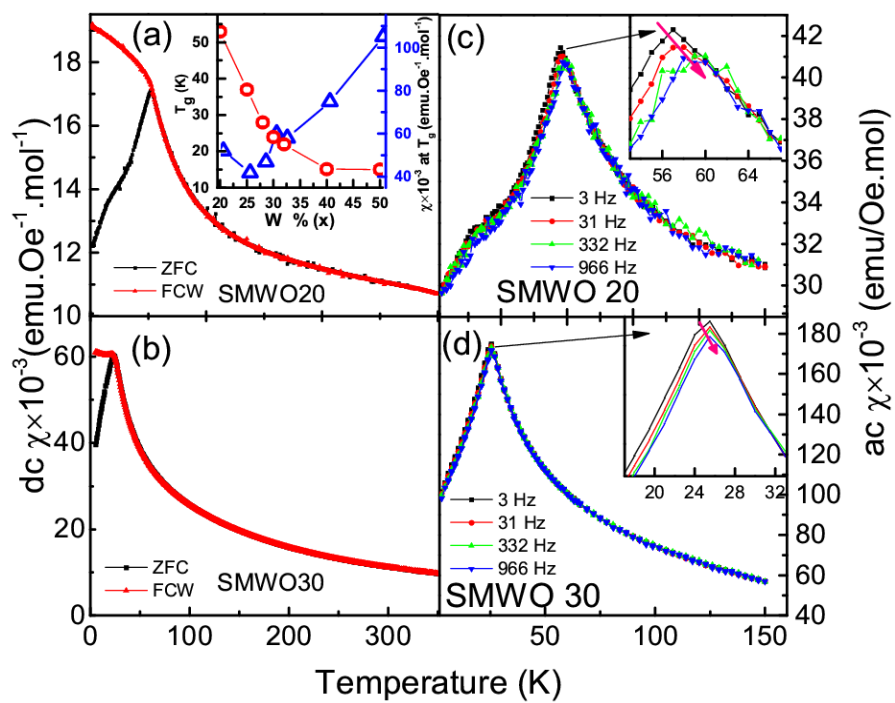

**Fig.2**

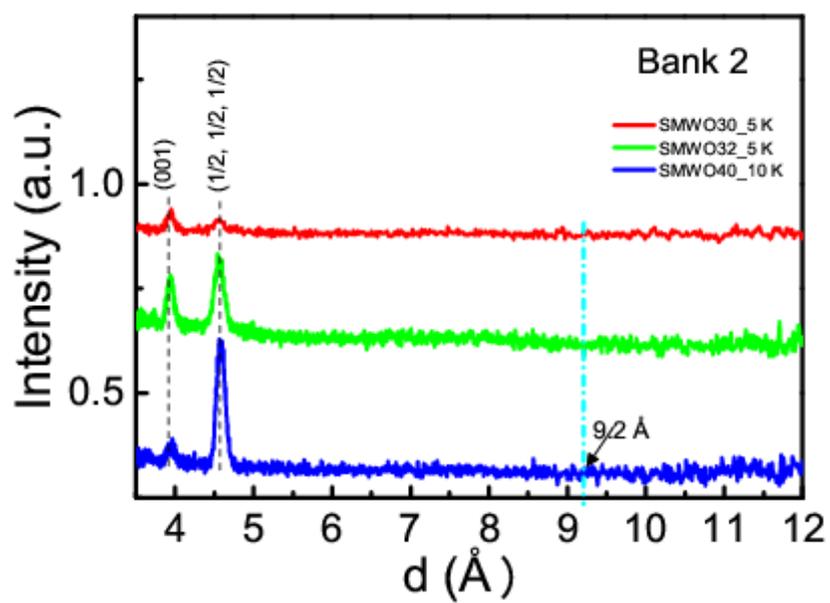

**Fig.3**

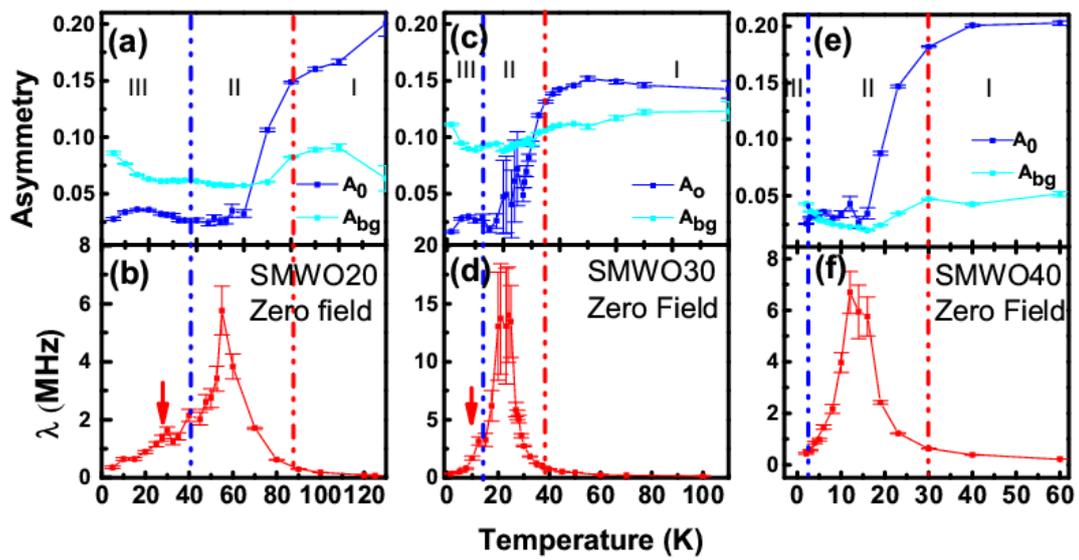

**Fig.4**

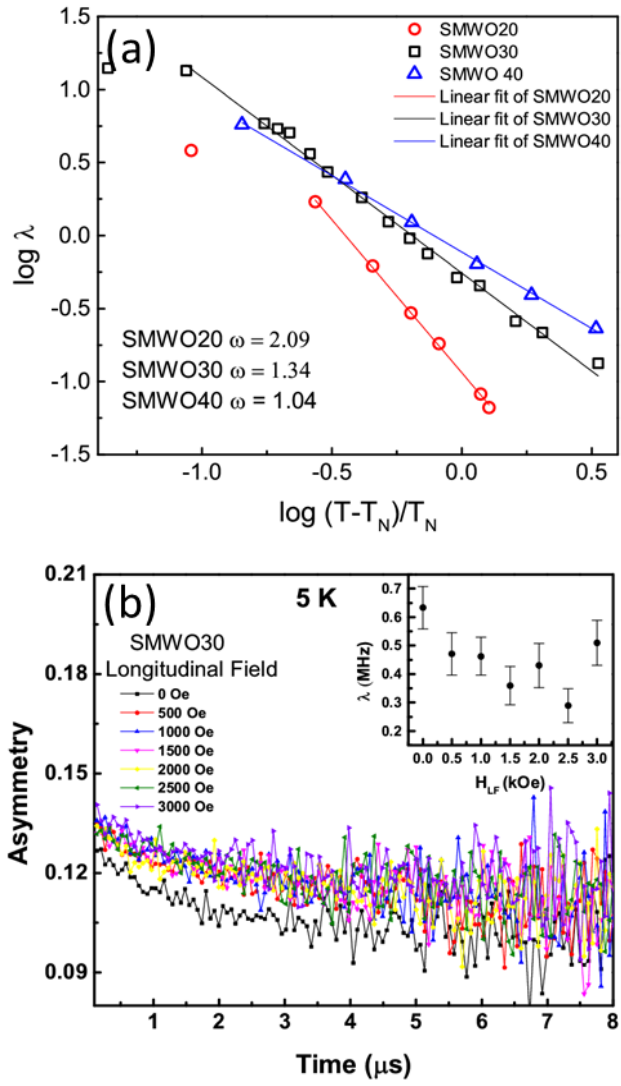

**Fig.5**

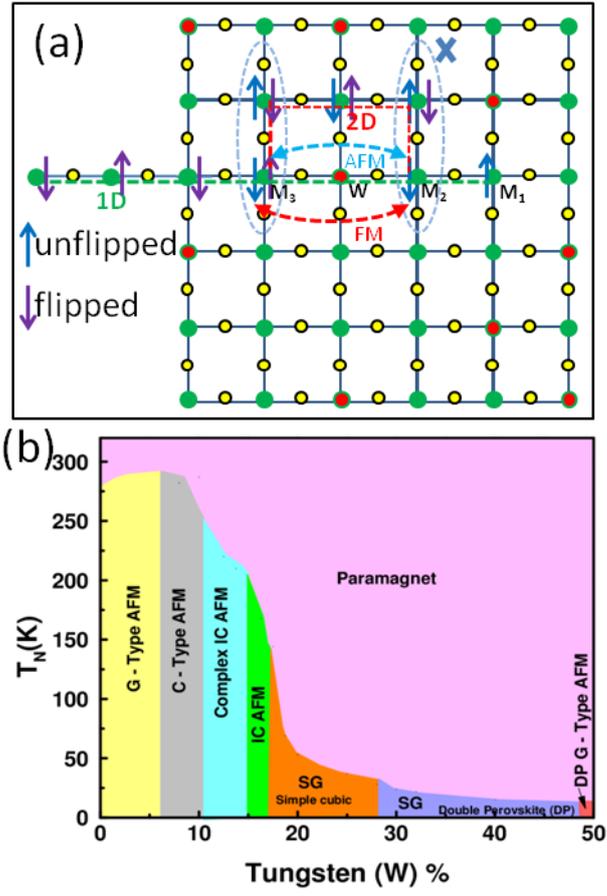

**Fig.6**